\documentclass[twocolumn]{aastex61hack}
\usepackage{natbib}
\newcommand{\HI}{\ensuremath{\mbox{\ion{H}{1}}}}
\newcommand{\HII}{\ensuremath{\mbox{\ion{H}{2}}}}
\newcommand{\OII}{\ensuremath{[\mbox{\ion{O}{2}}]}}
\newcommand{\OIII}{\ensuremath{[\mbox{\ion{O}{3}}]}}
\newcommand{\NII}{\ensuremath{[\mbox{\ion{N}{2}}]}}
\newcommand{\SII}{\ensuremath{[\mbox{\ion{S}{2}}]}}

\newcommand{\Halpha}{\ensuremath{\rm H \alpha}}
\newcommand{\Hbeta}{\ensuremath{\rm H \beta}}
\newcommand{\Hgamma}{\ensuremath{\rm H\gamma}}
\newcommand{\halpha}{H$\alpha$\relax}

\newcommand{\z}{$z$}

\newcommand{\percc}{cm$^{-3}$}

\newcommand{\kms}{km s$^{-1}$}

\newcommand{\ngc}{NGC~}

\newcommand{\lhalpha}{\ensuremath{L_{{\rm H}\alpha}}\relax}

\newcommand{\Rtwothree}{\ensuremath{R_{23}}}
\newcommand{\pymcz}{{\tt pyMCZ}}
\newcommand{\epsO}{\ensuremath{\epsilon ({\rm O})}}
\newcommand{\DeltaO}{\ensuremath{\Delta \epsilon ({\rm O})}}

\defcitealias{howk2018b}{Paper~II}

\begin{document}

\title{Extraplanar \HII\ Regions in Spiral Galaxies. I. Low-Metallicity Gas
Accreting through the Disk-Halo Interface of NGC 4013}

\author[0000-0002-2591-3792]{J. Christopher Howk}
\affiliation{Department of Physics, University of Notre Dame,
  Notre Dame, IN 46556, USA}
\affiliation{Instituto de Astrof\'{i}sica,
  Pontificia Universidad Cat\'{o}lica de Chile, Santiago, Chile}
\author{Katherine M. Rueff}
\affiliation{Department of Physics, University of Notre Dame,
  Notre Dame, IN 46556, USA}
\author{Nicolas Lehner}
\affiliation{Department of Physics, University of Notre Dame,
  Notre Dame, IN 46556, USA}
\author{Christopher B. Wotta}
\affiliation{Department of Physics, University of Notre Dame,
  Notre Dame, IN 46556, USA}
\author{Kevin Croxall}
\affiliation{Department of Astronomy, The Ohio State University,
  Columbus, OH 43210, USA}
\affiliation{Illumination Works LLC, 5550 Blazar Parkway \#150, Dublin,
   OH 43017, USA}
\author{Blair D. Savage}
\affiliation{Department of Astronomy, University of Wisconsin,
  Madison, Madison, WI 53706, USA}

\begin{abstract}

The interstellar thick disks of galaxies serve as the interface between the thin
star-forming disk, where feedback-driven outflows originate, and the distant
halo, the repository for accreted gas. We present optical emission line
spectroscopy of a luminous thick disk \HII\ region located at $z = 860$ pc above
the plane of the spiral galaxy \ngc 4013 taken with the Multi-Object Double
Spectrograph on the Large Binocular Telescope. This nebula, with an \halpha\
luminosity $\sim4-7$ times that of the Orion nebula, surrounds a luminous
cluster of young, hot stars that ionize the surrounding interstellar gas of the
thick disk, providing a measure of the properties of that gas.  We demonstrate
that strong emission line methods can provide accurate measures of relative
abundances between pairs of \HII\ regions. From our emission line spectroscopy,
we show that the metal content of the thick disk \HII\ region is a factor of
$\approx2$ lower than gas in \HII\ regions at the midplane of this galaxy (with
the relative abundance of O in the thick disk lower by $-0.32\pm0.09$ dex).
This implies incomplete mixing of material in the thick disk on small scales
(100s of parsecs) and that there is accretion of low-metallicity gas through the
thick disks of spirals. The inclusion of low-metallicity gas this close to the
plane of \ngc 4013 is reminiscent of the recently-proposed ``fountain-driven''
accretion models.

\end{abstract}


\section{Introduction}

The continued formation of stars in a massive, Milky Way-like galaxy requires
the addition of mass into the system from the intergalactic medium (IGM) to
replace that consumed by star formation \citep[e.g., ][]{kennicutt1983,
bauermeister2010, fraternali2012}. The timescales for IGM matter to reach a star
forming disk may be very long, and the processes by which it gets there are
quite varied. This is especially true for higher-mass galaxies, where any
accreted gas is expected to be heated as it interacts with the extended coronal
gas ($T \approx T_{\rm vir} \ga 10^6$ K, the virial temperature of the galaxy).
Once heated, the accreted gas is incorporated into the extended gaseous halo --
and hence unavailable for star formation -- until such time as it can cool and
be deposited into the central star forming regions. However, there are
significant questions about how effectively the halo of a Milky Way-like galaxy
can cool on its own \citep[i.e., the halo has a ``stability problem'' in that it
may be largely stable against cooling][]{binney2009}.

Recent models have invoked novel approaches to stimulating the cooling
of coronal gas, providing a potential new path to tapping this
extensive reservoir of matter to fuel star formation. These take the
form of ``feedback-induced'' or ``fountain-induced'' cooling
\citep{marinacci2010, marinacci2012, fraternali2017}. In these models,
metal-rich material ejected from the disk to large heights, $z$, above
the plane of a galaxy mixes with the coronal material. The mixing of
cold, metal-rich gas into the coronal matter provides the necessary
ingredients for enhanced cooling. Such a mechanism could also be at
play with material ejected through a ``galactic bore'' or hydraulic
jump near spiral arms \citep[e.g.,][]{martos1998} or perhaps even
through the stimulation of cooling by high velocity clouds
\citep{gritton2017}. Thus, the galactic fountain or other mechanisms
that eject disk material may be important to fueling the next
generation of star formation.

Disk galaxies often exhibit signatures of recently-ejected disk material from
either an active galactic fountain or other processes. A large fraction
\citep{howk1999} of nearby galaxies have thickened, rotating distributions of
gas and dust several kiloparsecs thick \citep[e.g., ][]{rand1990, dettmar1990,
howk2000, fraternali2001, heald2006, zschaechner2015, boettcher2016,
bizyaev2017} that can serve as boundary layers between the thin disks and hot
halos. These ``interstellar thick disks'' are complex -- with molecular,
neutral, and ionized gas at a broad range of temperatures detectable in emission
and absorption \citep{garcia-burillo1999, rueff2013, zschaechner2015} -- and
have extents characterized by exponential scale heights of of up to a few
kiloparsecs. Much of the matter making up interstellar thick disks may be
circulating through the first few kpc above the disk plane as part of a
``galactic fountain,'' carrying material upward for $\sim100$ Myr before it
returns \citep{shapiro1976, bregman1980}, or perhaps ejected as part of a
hydraulic jump caused by shocks as gas falls into the potential well of a spiral
arm \citep{martos1998}. This metal-rich disk material may supply the conditions
needed to spur the cooling of coronal material. However, to date there has not
been a straightforward path to testing for the presence of cooled coronal matter
in the thick disk. Some studies have argued the rotational lag of the thick disk
gas may indicate the presence of infalling gas \citep[][not necessarily from the
corona]{fraternali2006}, although there may be other explanations for the lag
\citep{struck2009}. Dust-to-gas indicators may provide a method of assessing the
contribution from infalling matter to the thick disk \citep{peek2009a,
howk2012b}.

As part of their studies of gas in the ``disk-halo interface\footnote{We
typically use the terms thick disk and disk-halo interface interchangeably to
describe the interstellar material within the first few kpc above that largely
co-rotates with the underlying disk of a galaxy.},'' several groups have noted
the presence of extraplanar \HII\ regions far from the planes of edge-on
galaxies \citep{walterbos1991, howk2000, tullmann2003, rueff2013, stein2017}. In
some cases these include stars formed {\em in situ} \citep{howk2018b}, likely
from the dense gas that is often seen threading the thick disk \citep{howk1999,
howk2000, rueff2013}.  No matter where they formed, the massive stars ionize and
energize the surrounding thick disk gas to produce these extraplanar nebulae.
The emission lines from these \HII\ regions can be used as a probe of the gas in
the thick disk, allowing us to study its metallicity \citep[e.g.,
][]{tullmann2003, stein2017}. The gas-phase abundance or metallicity of thick
disk gas can be an indicator of its origins: gas dominated by outflows from the
disk will have disk-like metallicities, while material accreting from the corona
or even directly from the IGM will have a much lower metal content. Thus, these
thick disk nebulae allow us to ask fundamental questions about the nature of the
baryon cycle in spiral galaxies.

There are a few existing measurements of abundances in thick disk \HII\ regions,
with a mix of results for their abundances relative to the disks of their host
galaxies. \cite{tullmann2003} were the first to present convincing spectroscopy
of extraplanar \HII\ regions, in this case in the dwarf galaxy \ngc 55. The two
extraplanar \HII\ regions in this galaxy (found to heights approaching $z
\approx 2$ kpc) are somewhat more enriched than the disk gas.\footnote{This
conclusion is different than the original conclusions of \cite{tullmann2003},
who were hampered by the diagnotics available at the time.  See the details in
\S \ref{sec:vertgradients}.} Recently, \cite{stein2017} have provided new
measurements of abundances in a total of three extraplanar \HII\ regions in the
galaxies \ngc 3628 and \ngc 4522 (at distances of $1.4 \la z \la 3.0$ kpc from
the midplane). These nebulae have abundances mostly consistent with their host
galaxies, as expected given their likely origin in the tidally- or ram
pressure-stripped material seen in \HI\ maps of these galaxies
\citep{wilding1993,kenney2004}.  These are thus reminiscent of the \HII\ regions
in the outer disk / tidal material studied by \cite{werk2011}.

This is the first in a series of papers discussing the nature of extraplanar
\HII\ regions in edge-on spiral galaxies. Here we focus on abundance
measurements in an extraplanar \HII\ region located firmly within the
interstellar thick disk of the edge-on galaxy \ngc 4013. This nebula was briefly
discussed by \cite{rueff2013} as part of our study of the thick disk
interstellar medium (ISM). It lies at a height of $z = 860$ pc above the plane,
within an interstellar thick disk that seems to be strongly influenced by an
active galactic fountain \citep{howk1999, rueff2013}. This region of the thick
disk shows \HI\ 21-cm emission from a warm neutral medium
\citep{zschaechner2015}, \Halpha\ emission from the diffuse ionized gas
\citep[DIG; ][]{rand1996, rueff2013}, as well as CO emission
\citep{garcia-burillo1999} and dust absorption \citep{howk1999, rueff2013} from
a cold neutral medium. Thus, the placement of this nebula allows us to probe the
mixing of metals in a region typical of fountain-fed thick disks in spiral
galaxies. A companion paper, \cite{howk2018b} (hereafter
\citetalias{howk2018b}), describes the stars responsible for ionizing the nebula
and the physical ingredients required to explain their {\em in situ} formation
in the thick disk.

We discuss our emission line spectroscopy of the extraplanar \HII\ region in \S
\ref{sec:spectroscopy}, including our observations, data reduction, and emission
line intensities. We use these observations to assess the abundance of this
extraplanar nebula relative to that of gas in the thin disk in \S
\ref{sec:metallicity}, including a detailed discussion of our approach to
estimating the relative abundance. In \S \ref{sec:discussion} we discuss our
derived metallicity offset for the thick disk \HII\ region and the implications
for mixing in \ngc 4013. We discuss this measurement in the context of existing
measurements from extraplanar \HII\ regions and Milky Way halo clouds in \S
\ref{sec:vertgradients}, and we summarize our principal conclusions in \S
\ref{sec:summary}.  Throughout we assume a distance $D=17.1\pm1.7$ Mpc to \ngc
4013 (equivalent to a distance modulus $31.17\pm0.10$ mag). This is the mean
distance to the galaxy group with which \ngc 4013 is associated, group 102 from
the catalog of \cite{tully2008}.

\section{Emission Line Spectroscopy of the Extraplanar Nebula
  NGC~4013 EHR1}
\label{sec:spectroscopy}

\begin{figure*}
\plotone{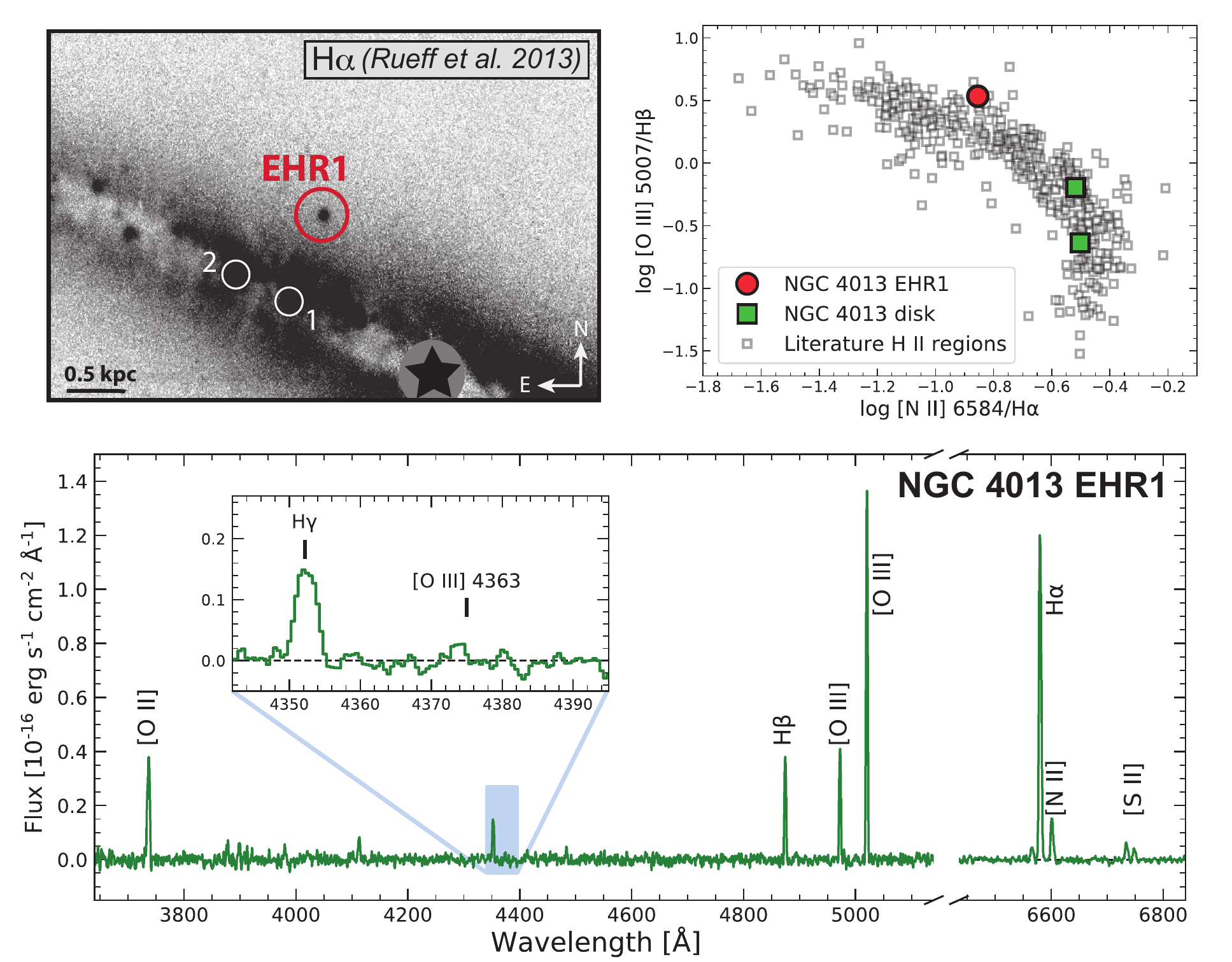}

\caption{Montage of results on \ngc 4013 EHR1.{\em Top left:} WIYN
  \Halpha\ image of the edge-on spiral galaxy NGC 4013
  \cite{rueff2013} showing the location of the extraplanar
  \HII\ region EHR1 as well as the two disk \HII\ regions used to
  define the disk reference abundance. \ngc 4013 EHR1 is seen at a
  height $z = 860$ pc. (The black star hides mis-subtracted
  emission about a very bright foreground Milky Way star.)  {\em Top
    right:} Observed emission line ratios for EHR1 compared with
  literature measurements of \HII\ regions in the disks of spiral
  galaxies \citep{berg2015, croxall2015, croxall2016,
    bresolin2009}. These ratios for EHR1 are similar to those seen in
  normal \HII\ regions. Combined with its velocity, which is
  consistent with the diffuse ionized gas emission in this region of
  \ngc 4013, this identifies EHR1 as an \HII\ region in the thick disk
  of this galaxy.  {\em Bottom:} Partial LBT/MODS1 spectrum of EHR1
  with several of the strongest lines identified. The inset shows the
  spectral region encompassing H$\gamma$ and the undetected auroral
  \OIII\ 4363 \AA\ transition (the shaded blue box shows this region
  in the larger spectrum). \label{fig:montage}}
\end{figure*}

We take advantage of an extraplanar \HII\ region (EHR) to probe the origins of
gas in the thick disk of the edge-on spiral galaxy \ngc 4013.  The upper left
panel of Figure \ref{fig:montage} shows an \Halpha\ image of part of \ngc 4013
from \cite{rueff2013}. This nebula -- \ngc 4013 EHR1 -- is marked by the red
circle. EHR1 has a luminosity $\sim4 - 7$ times that of the Orion nebula at a
projected height above the midplane $z = 860$ pc and projected radial distance
of $R\sim2.5$ kpc from the galaxy's center.  This section describes the
properties of our low-resolution optical long-slit spectra of \ngc 4013 EHR1 and
two reference regions in the thin disk using one of the Multi-Object Dual
Spectrographs (MODS) on the Large Binocular Telescope (LBT).  The final
extracted spectrum of EHR1 is shown in the bottom panel of Figure
\ref{fig:montage}, and select emission line ratios for \ngc 4013 EHR1 (and the
two disk regions) derived from these data are compared with a sample of
extragalactic \HII\ regions from the local universe \citep{bresolin2009,
berg2015, croxall2015, croxall2016} in the upper right panel.  The emission line
spectrum of EHR1 is characteristic of \HII\ regions found in the disks of spiral
galaxies.  This and its radial velocity, which is consistent with that of the
local diffuse ionized gas, demonstrate that EHR1 is an \HII\ region in the thick
disk of \ngc 4013.

%%%%%%%%%%%%%%%%%%%%%%%%%
%%
%% Figure 1

\subsection{Observations and Basic Reduction}

We used the first of the MODS instruments (MODS1) installed on the LBT to obtain
spectra of both \ngc 4013 EHR1 and the two reference \HII\ regions in the disk
of \ngc 4013 (Figure \ref{fig:montage}). The coordinates, projected radial
distance from the galaxy center, $R$, and $z$-heights of the three nebulae are
listed in Table \ref{tab:HIIregions}. The MODS instruments are described by
\cite{pogge2010}, and examples of MODS1 spectroscopy applied to \HII\ region
abundances can be found in the CHAOS papers \citep{berg2015, croxall2015,
croxall2016}.

MODS1 is a double spectrograph with separate blue and red
channels. Both channels are used simultaneously, and the light is
split by a dichroic with a crossover at $\lambda \sim 5500$ \AA. We
adopted the standard set-up of the instrument, with the G400L grating
(400 lines mm$^{-1}$) in the blue and the G670L (250 lines mm$^{-1}$)
grating in the red. In this configuration, MODS provides wavelength
coverage over $3200 \la \lambda \la 10000$ \AA, which is sampled after
reduction at 0.5 \AA\ per pixel. All of our observations used the
1\arcsec -wide longslit mask, which is a sequence of five slits of length 1\arcmin\ aligned end-to-end with a few arcsec gaps between them. This has
no practical effect for our work compared with a contiguous longslit,
and we will discuss this mask as if it were a traditional single
longslit. With our 1\arcsec\ slit width, both spectrograph channels
give a spectral resolution $R \sim 1200$.

We observed \ngc 4013 EHR1 on 2012 June 15 and the disk \HII\ regions on 2015
January 20. In the case of EHR1 the slit was aligned at the parallactic angle,
and the object observed for $3\times20$ minutes. For the disk observations, the
slit was placed along the disk of \ngc 4013 (at PA $64\fdg2$ E of N), which was
within 10$^\circ$ of the parallactic angle during those observations. The disk
was observed for $3\times20$ minutes. The seeing in both cases was under
1\arcsec\ and both observations were obtained in photometric conditions.

\begin{deluxetable}{lllcl}
  \tablewidth{0pc}
\tablecaption{Observed NGC 40\ \HII\ Regions \label{tab:HIIregions}}
\tablehead{\colhead{Name} &
    \colhead{RA (J2000)} & \colhead{Dec (J2000)}
    & \colhead{$R$ (kpc)} & \colhead{$z$ (pc)}}
\startdata
  EHR1   & 11 58 33.2 & +43 57 11.85 & 2.5 & 860 \\
  disk 1 & 11 58 33.7 & +43 57 00.86 & 2.3 & $\approx0$ \\
  disk 2 & 11 58 34.2 & +43 57 03.88 & 2.9 & $\approx0$ \\
\enddata
\end{deluxetable}

\subsection{Data Reduction and Spectral Extraction}

The basic reduction and spectral extraction is done using the MODS reduction
pipeline\footnote{The MODS reduction pipeline was developed by Kevin Croxall
with funding from NSF Grant AST-1108693. Details can be found at
\url{http://www. astronomy.ohio-state.edu/MODS/Software/modsIDL/}.} developed by
K. Croxall and based on the XIDL code\footnote{The XIDL code of J.X. Prochaska
is available at \url{https://github.com/profxj/xidl}.}. The code has been
applied and briefly described as part of the CHAOS project
\citep{berg2015,croxall2015,croxall2016}, and we refer the reader particularly
to \cite{berg2015} for the most detailed description.  Following those works we
assume a 2\% uncertainty in the flux calibration, adding this in quadrature to
the errors associated with general Poisson noise, read noise (negligible), sky
and background subtraction uncertainties.

We extracted the spectrum for \ngc 4013 EHR1 using a $1\farcs5$ extraction box,
while we used a $2\farcs0$ box for the disk \HII\ regions in order to account
for all of the flux.  In each case we extracted equally-sized background spectra
on either side of the objects of interest. One of the key components of the data
reduction is assessing and removing the background from the emission line
spectra of the \HII\ regions. This is particularly important for EHR1, for which
the long-slit spectra include contributions from the nebula as well as from both
from the old stellar population of the thick disk and bulge (continuum) plus the
diffuse ionized gas (DIG) of the thick disk (emission lines). The DIG emission
line spectrum is quite different from the nebular emission from EHR1 itself,
with ratios \NII /\Halpha $\, \approx 1$, for example, compared with a ratio of
$\approx 0.13$ for the \HII\ region. Thus proper background subtraction is
important, as while the \Halpha\ emission from the DIG is much fainter (only
$\sim13\%$) than that of EHR1, the forbidden line strengths can contribute
significantly to the uncorrected spectrum (with DIG lines matching those from
EHR1).

\begin{figure*}
\plotone{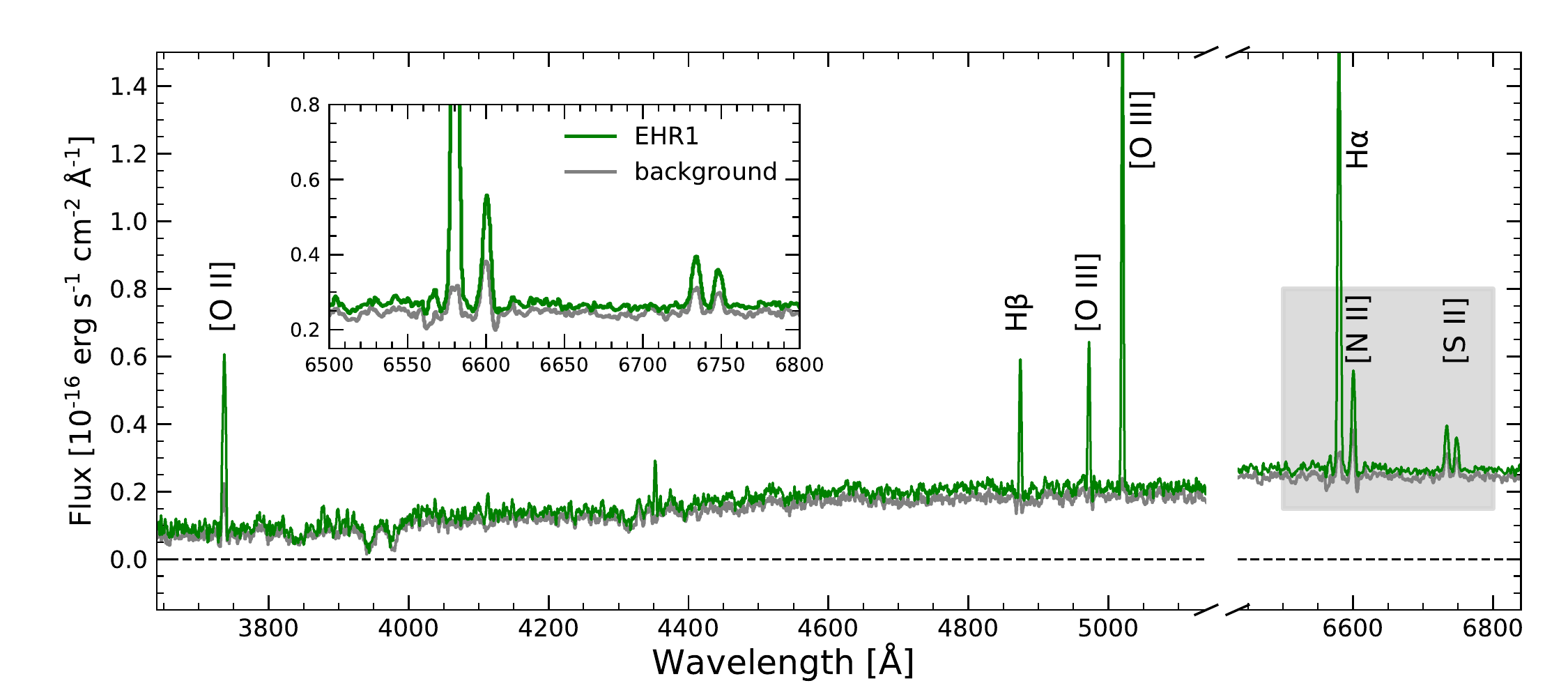}
\caption{A comparison of the spectrum extracted toward EHR1 (green) with our
estimate of the background light (grey) based on a combination of spectra
extracted on either side of our target. The spectrum toward EHR1 includes line
emission from the nebula EHR1 itself, line emission from the DIG, and continuum
emission from the bulge and thick disk light in this direction. The background
spectrum is difficult to see in most of the spectrum given the good agreement
between the continuum emission in the target and background spectra. The stellar
continuum emission from the stars powering EHR1 itself is not expected to be
detected \citepalias[it has $V \sim 23.7$ mag; see][]{howk2018b}, nor is it
seen. The strong absorption lines seen in both the background and EHR1
directions are from the old stellar populations in these directions; \ion{Ca}{2}
$\lambda, \lambda$3933, 3968 are the most prominent.  The inset shows the region
around \halpha, including the forbidden \NII\ and \SII\ transitions, in order to
demonstrate the differing contributions the background DIG makes to these lines.
A slight residual flux difference is seen between the object and background
spectra, which we remove by subtracting a low-order polynomial fit to the
residual continuum. \label{fig:bgspectrum}}
\end{figure*}

We have tested the background subtraction for EHR1 in two ways; both
give indistinguishable results. In the first approach, we assume a
fully-empirical background, using the geometric mean of the spectra from regions
immediately adjacent to the extraction box for EHR1. These spectra were
extracted just as those of EHR1. We use a geometric mean under the assumption
that the DIG and stellar background light have exponential decays with height.
As shown in Figure \ref{fig:bgspectrum}, the resulting background spectrum is a
very good match to the shape and intensity of the continuum light underlying
EHR1. There is a slight residual continuum that we remove with a low-order
polynomial (the residuals are roughly flat and $\la 10\%$ of the local flux).
This may arise from slightly different extinction or stellar populations, or if
the use of the geometric mean of the background regions results in a slight
departure from the true stellar populations underlying EHR1.

The second approach uses a full spectral fit of both the EHR1 spectrum and
seperately the geometric mean of the two background region spectra (on either
side of the nebula) using the penalized pixel fitting (pPXF) software of
\cite{cappellari2017}. In this approach, we fit composite stellar models and
nebular emission lines to the spectrum over the wavelength range $3500 \la
\lambda \la 7500$ \AA. The wavelength range is set by the useful range of the
MILES spectral libraries \citep{sanchez-blazquez2006} that we adopt to construct
the composite spectrum. While we fit the emission lines as part of the process,
this is done only to allow for their affects on the fitted background stellar
spectra. We subtract the pPXF-derived stellar flux models from the EHR1 and
background spectra, leaving a pure emission line spectrum from the nebula and
DIG. We then subtract the background DIG emission line spectrum from that observed
toward EHR1.

These two approaches give final emission line spectra for EHR1 that
are indistinguishable. We adopt the direct empirical background moving
forward. The background assessment for the two disk \HII\ regions
followed a similar approach.

\subsection{Emission Line Measurements}

We measured the intensities of the emission lines by fitting the full
spectrum with a Gaussian emission line model, returning the
velocities, velocity dispersions, and intensity for all potential
lines of interest.  We generally forced the dispersions of lines from
the same ion to be the same. Within the same spectrograph channel, the
central velocities of the collisionally-excited metal lines were tied
in the fitting process, as were those of the Balmer lines.  Once the
emission lines were fitted, we also use the central velocities and
dispersions to measure emission line intensities via direct
integration of the lines.

For \ngc 4013 EHR1 and the reference region disk 1, the line profiles were well
fit by our assumed Gaussian models. The same is not true for the disk 2 region,
which shows substructure in the lines associated with the structure in the
nebular emission as projected onto the slit. There is also some evidence that
slit losses for the disk 2 spectrum may be more significant in the red than the
blue, as the the line shapes are not in great agreement between the
spectrographs, and there is a slight velocity offset between the two channels
that is not seen in the disk 1 spectra. This is also reflected in the comparison
of the integrated and fitted line intensities, which are quite similar for EHR1
and disk 1, but have larger discrepancies in the disk 2 spectrum (seemingly due
to the non-Gaussian line profiles in the disk 2 spectrum). We feel the
integrated lines are more representative of the true fluxes for at least the
reference disk 2 region, and we adopt those throughout. Given the peculiarities
of the disk 2 spectra, as well as the coincidence in projected radial distance
between EHR1 and disk 1, we favor the results from disk 1 and will use this
object as our main reference against which to judge the properties of EHR1. We
note that the velocities of the reference disk 1 ($v_{\rm disk\ 1} = 698\pm25$
\kms\ heliocentric) and EHR1 ($v_{\rm disk\ 1} = 720\pm25$ \kms) are also
similar, which could indicate they arise from similar locations in the galaxy.

We assess the foreground extinction toward each nebula by comparing the ratio of
\Hbeta\ to \Hgamma\ assuming $T_e \approx 10^4$ K. We do this rather than the
more typical \Halpha /\Hbeta\ ratio in order to avoid comparisons across the
dichroic break. For disk 2, in particular, we subsequently adjust the scale of
the red spectra such that \Halpha /\Hbeta\ is in agreement with expectations. We
have experimented with using the \Halpha /\Hbeta\ ratio to determine the
extinction, which always leads to larger extinction values; however, our
abundance results are insensitive to this choice, as the emission line ratios we
use to derive abundances typically use line pairs close in wavelength
(\ref{sec:SELscales}).

The resulting extinction-corrected line intensities relative to \Hbeta\ and the
adopted color excesses, $E(B-V)$, for each nebula are given in Table
\ref{tab:intensities}. We expect that the Milky Way contributes $<0.02$ mag to
these color excesses \citep{schlafly2011}.

%%%%%%%%%%%%%%%%%%%%%%%%%%%%%%%%%%%%%%%%%%%%%%%%%%%%%%%%%%%%%%%%%%%%%%
%\clearpage
\begin{deluxetable}{lccc}
  \tabletypesize{\scriptsize}
  \tablewidth{0pc}
  \tablecolumns{4}
  \tablecaption{Extinction-corrected Emission Line Intensities \label{tab:intensities}}
  \tablehead{\colhead{} & \colhead{}  &
    \colhead{$100 \times I_\lambda / I_{{\rm H}\beta}$} & \\
    \cline{2-4}\\
    \colhead{Line} & \colhead{EHR1} & \colhead{disk 1} & \colhead{disk 2}}
  \startdata
  \OII\ 3727 & $226.9 \pm 6.9$ & $254.2 \pm 10.2$ & $277.8 \pm 5.6$ \\
  \Hgamma\ 4341 & $47.0 \pm 2.2$ & $46.6 \pm 2.7$ & $46.6 \pm 1.0$ \\
  \OIII\ 4364 & $< 6.1$ & $< 6.9$ & $< 1.6$ \\
  \Hbeta\ 4861 & $\equiv 100.0$ & $\equiv 100.0$ & $\equiv 100.0$ \\
  \OIII\ 4960 & $107.3 \pm 2.7$ & $6.3 \pm 1.5$ & $19.6 \pm 0.5$ \\
  \OIII\ 5008 & $342.3 \pm 7.1$ & $24.0 \pm 1.6$ & $63.5 \pm 1.3$ \\
  \NII\ 6549 & $12.3 \pm 0.9$ & $29.3 \pm 0.7$ & $29.6 \pm 0.6$ \\
  \Halpha\ 6564 & $289.6 \pm 5.9$ & $285.8 \pm 5.7$ & $286.8 \pm 5.7$ \\
  \NII\ 6585 & $40.5 \pm 1.3$ & $91.6 \pm 1.9$ & $86.7 \pm 1.7$ \\
  \SII\ 6718 & $16.3 \pm 1.0$ & $35.9 \pm 0.8$ & $40.9 \pm 0.8$ \\
  \SII\ 6732 & $9.0 \pm 1.0$ & $25.0 \pm 0.6$ & $29.0 \pm 0.6$ \\
  \hline
  $E(B-V)$ (mag) & $0.22\pm0.07$  & $0.42\pm0.09$ & $0.41\pm0.04$ \\
  \enddata
  \tablecomments{Upper limits are 3$\sigma$. The intensity reported for
    the \OII\ 3727 line is the sum of the two members of the \OII\
    doublet.}
\end{deluxetable}

\subsection{Physical Conditions of the Thick Disk \HII\ Region}

Our spectra provide some constraints on the physical properties of the thick
disk \HII\ region \ngc 4013 EHR1. For none of the three \HII\ regions have we
detected the weak temperature-sensitive auroral lines needed to derive their
temperatures (e.g, \OIII\ $\lambda$4363 shown in the inset of Figure
\ref{fig:montage}). Our non-detection of \OIII\ $\lambda$4363 for EHR1 provides
an electron temperature upper limit $T_e(\mbox{\OIII}) \la 13,200$ K
($3\sigma$). Many studies have found the  temperatures and gas-phase oxygen
abundances of \HII\ regions are correlated; our temperature limit is
characteristic of \HII\ regions with $\epsO \equiv 12+\log ({\rm O/H}) \ga 7.7$,
assuming the correlation of \cite{lopez-sanchez2012}. This lower limit is well
below the estimates of \epsO\ for EHR1 we derive below (\S
\ref{sec:montecarlo}). Thus we are likely quite far from detecting the \OIII\
$\lambda$4363 emission.

The ratio \SII\ $\lambda$6717/$\lambda$6731 is consistent with the low-density
limit of this diagnostic, placing an upper density limit $ n_e \le 100$ \percc.
The \Halpha\ luminosity of the nebula, $\lhalpha \ge (4.0\pm1.2) \times10^{37}$
ergs s$^{-1}$, requires an H-ionizing photon production rate $\log Q_{0} \ge
49.4$ \citepalias[units of photons s$^{-1}$;][]{howk2018b}. Thus EHR1 is a
relatively luminous H II region, with an \halpha\ luminosity $\lhalpha \sim 4-7$
times that of the Orion nebula\footnote{We assess the Orion nebula luminosity
based on the distance from \citet{sandstrom2007} and integrated flux from
\citet{rumstay1984}.}. The ionizing photon flux required to power this is
equivalent to that of $\sim6$ O7 V stars \citepalias[see][]{howk2018b}.  The
\Halpha\ emission is unresolved in our images \citep[$R_S \la 33$ pc;
][]{rueff2013}. Using this radius, we derive a lower density limit assuming the
nebula is a Str\"{o}mgren sphere, providing overall limits of $5 \le n_e \le
100$ \percc.

\section{Relative Metallicity of the Thick Disk \HII\ Region}
\label{sec:metallicity}

Given our non-detection of the auroral lines, we cannot derive ``direct''
gas-phase abundance estimates from our data. We must rely on the so-called
``strong emission line'' (SEL) techniques to estimate the gas-phase oxygen
abundance. There are a variety of SEL scales available, which we take to include
a method (e.g., the ``O3N2 method'') and a specific calibration of that method
\citep[e.g., the calibration of the N2 method by][]{pettini2004}. There can be
significant systematic errors in SEL-based absolute metallicities, $\epsO \equiv
12+\log ({\rm O/H})$, for a given \HII\ region, and the errors are different for
each chosen scale. However, the {\em relative} metallicities of two \HII\
regions, $\DeltaO \equiv \epsO_{\rm j} - \epsO_{\rm k}$, within a given scale
are robust, as we demonstrate below \citep[see also ][]{kewley2008}.  For this
reason, we focus on calculating the relative abundance of the \HII\ region EHR1
compared with our reference nebulae in the disk. Thus, we are focusing on
calculating abundance gradients or offsets rather than the absolute abundances.

\subsection{Choosing SEL Abundance Scales}
\label{sec:SELscales}

There are a multitude of SEL scales available in the
literature. Individual scales, even using the same method, can give
different absolute abundances \citep{kewley2008, curti2017}.  In this
work, we make use of a limited set of abundance scales for determining
the abundance offset between EHR1 and the disk of \ngc 4013. We choose
a set of scales based on their applicability to the current situation,
focusing on those calibrated in the metallicity regime describing our
objects (which all scales place in the range $\epsO \approx 8.25$ to
8.80). We exclude from consideration those scales that are
double-valued with turn-around points that are nearly coincident with
the metallicities at which we appear to be working \citep[e.g., the
  $R_{23}$ method; ][]{kewley2008}.

In some cases, several calibrations of a given metallicity method (indicator)
give good results that could be useable. However, we use only one calibration
for each method in deriving our final results in order to minimize potentially
correlated systematic issues across the calibrations. We have also avoided
methods that rely on the \OII\ 3727 lines given the large uncertainties in our
extinction correction (although their adoption would not lead to a different
result).

To choose which are the most appropriate scales for our purposes, we have
assessed which scales provide robust measures of abundance offsets, \DeltaO,
analyzing 17 separate scales from the literature. For each scale we calculate
\DeltaO\ from direct method abundances for a sample of literature \HII\ regions
in spiral galaxies \citep{bresolin2009, berg2015, croxall2015, croxall2016} and
compare them with calculated offsets from each SEL scale. We calculate
\DeltaO$_{SEL}$ for a sample of ``low metallicity'' \HII\ regions (with direct
abundances $8.20 \le \epsO \le 8.50$) compared with a ``high metallicity''
sample (with $8.50 \le \epsO \le 8.80$).  These metallicity regimes bracket the
abundance ranges predicted in an initial determination of the \HII\ regions in
\ngc 4013.  We compare the implied offsets derived using the SEL abundances with
those from the direct methods to identify which SEL methods and calibrations
give reliable results.  We also assess the dispersion in the SEL offsets about
the direct method results. In Table \ref{tab:SELscales} we summarize the median
discrepancies in \DeltaO\ when comparing the SEL and direct results as well as
the intrinsic dispersions for the scales we considered.

\begin{deluxetable}{cccc}[ht]
  \tabletypesize{\scriptsize}
  \tablewidth{0pc}
  \tablecolumns{4}
  \tablecaption{Characteristics of SEL Metallicity Scales \label{tab:SELscales}}
  \tablehead{\colhead{Method} &
    \colhead{Calibration\tablenotemark{a}} &
    \colhead{$\DeltaO_{\rm SEL} - \DeltaO_{\rm Direct}$} & \colhead{$\sigma_{\DeltaO}$\tablenotemark{b}}}
  \startdata
  \multicolumn{4}{c}{Adopted scales for calculating \DeltaO} \\
  \hline
  N2   & PP04  & $+0.024$ & 0.169 \\
  O3N2 & M13   & $+0.010$ & 0.155 \\
  O3   & C17   & $-0.009$ & 0.178 \\
  \hline
  \multicolumn{4}{c}{Reliable scales for calculating \DeltaO} \\
  \hline
  N2   & M13    & $+0.030$ & 0.136 \\
  N2   & C17    & $+0.002$ & 0.148 \\
  N2O2 & PMC09  & $-0.012$ & 0.217 \\
  O3N2 & C17    & $-0.024$ & 0.176 \\
  O3O2 & C17    & $+0.054$ & 0.196 \\
  \hline
  \multicolumn{4}{c}{Less-reliable scales for calculating \DeltaO} \\
  \hline
  N2 &  KK04 &  $-0.016$ & 0.216 \\
  N2 &   M08 &  $-0.110$ & 0.251 \\
  O3N2 &  PP04 &  $-0.108$ & 0.229 \\
  O3N2 &   M08 &  $-0.188$ & 0.349 \\
  O3N2 & PMC09 &  $-0.096$ & 0.218 \\
  O3O2 &   M08 &  $-0.081$ & 0.359 \\
  O3 &   M08 &  $-0.158$ & 0.311 \\
  N2O2 &  B07  &  $-0.070$ & 0.169 \\
  N2O2 &  KD02  &  $-0.046$ & 0.244 \\
  \enddata
  \tablenotetext{a}{Calibration scale references -- B07
    \citep{bresolin2007}, C17 \citep{curti2017}, KD02
    \citep{kewley2002}, KK04 \citep{kobulnicky2004}, M08
    \citep{maiolino2008}, M13 \citep{marino2013}, PP04
    \citep{pettini2004}, PMC09 \citep{perez-montero2009}}
  \tablenotetext{b}{Dispersion in the $\DeltaO_{\rm SEL} -
    \DeltaO_{\rm Direct} $ in dex.}
\end{deluxetable}

Based on these results, we adopt the following SEL scales: the N2
scale from \cite{pettini2004}, the O3N2 scale of \cite{marino2013},
and the O3 scale of \cite{curti2017}.  All three adopted scales
provide have mean values of $| \DeltaO_{\rm SEL} - \DeltaO_{\rm
  Direct} | \la 0.03$ dex and dispersions in this quantity $\approx
0.15$ to 0.18 dex. Table \ref{tab:SELscales} gives the results of the
above analysis for all 17 scales we considered.  We break these down,
showing the three scales we ultimately adopt, the scales we did not
adopt adopt but still considered reliable for such calculations, and
those that were not appropriate for our uses. Our assessment of the
appropriateness of the scales is based on measures beyond simply this
table. For example, we rejected some of the scales because they show
non-linear slopes in the relationship between the SEL and direct
method abundances, with breaks in the relationships that do not bias
too much the mean behavior of the indicators but might for the study
of individual objects. We note that the methods that rely on
\OII\ tend to have large dispersions and \DeltaO\ offsets generally.

An important take-away from our analysis is the SEL-based calculations of
\DeltaO\ are in fact quite consistent with one another and with the direct
method results when using any of the scales deemed reliable in Table
\ref{tab:SELscales}. The SEL and direct method calculations of \DeltaO\ are, as
expected, much more consistent than the results for the absolute abundances,
\epsO. The absolute abundances, \epsO, can have systematic offsets of 0.1 -- 0.3
dex in this range of metallicities, while the discrepancies in the relative
abundance offsets, \DeltaO, are more typically in the 0.0 -- 0.1 dex range.
Although we will apply a limited number of abundance scales to our
determination of \DeltaO, our final results are not particularly sensitive to
the specific choice of scales from among those deemed reliable in Table \ref{tab:SELscales}.

\subsection{Estimating the Relative Abundance \DeltaO}
\label{sec:montecarlo}

We use a Monte Carlo approach based on the \pymcz\ code
\citep{bianco2016} to assess the posterior distribution functions of
the abundances $\epsO \equiv 12+\log ({\rm O/H})$ for each
\HII\ region and subsequently of $\DeltaO \equiv \epsO_{\rm EHR1} -
\epsO_{\rm ref}$ between EHR1 and the reference regions. This code
uses Monte Carlo sampling of the input emission line intensity
distributions to assess the posterior distribution functions of SEL
abundances from a choice of abundance scales.  In our approach, we
sample the emission line intensities for a given \HII\ region 4000
times (assuming a normal distribution with central values and
dispersions given from measurements), calculating the extinction and
gas-phase abundance implied by each sample. The net result is a
posterior distribution of gas-phase oxygen abundances for each
\HII\ region from each SEL scale.

We consider the relative abundances of EHR1 with two disk
\HII\ regions shown in Figure \ref{fig:montage}, which we take to be
representative of the disk as a whole. We calculate a posterior
distribution for \DeltaO\ in each scale by comparing the Monte Carlo
samples of EHR1 with the randomly-ordered samples for the reference
disk regions. An individual \HII\ region will not lie directly on the
calibrated relationship between SEL ratios and gas-phase abundance due
both to errors in the emission line intensities, but also due to
intrinsic differences in physical conditions and ionization states
that are not accounted for in the SEL methodology.  We account for
this intrinsic scatter of individual \HII\ regions about the mean SEL
scales by adding normally-distributed random offsets when calculating
\DeltaO\ for each pair of Monte Carlo samples. The dispersions adopted
for each SEL scale are drawn from Table \ref{tab:SELscales}. This
source of uncertainty (typically $\sim0.15$ to 0.20 dex for reliable
SEL scales) dominates over the propagated observational uncertainties
($\sim0.02$ to 0.04 dex). In addition, for each SEL scale we add a
correction to the \DeltaO\ values to account for the offsets between
\DeltaO\ derived from that scale compared with those derived from
direct abundance measurements. We use the corrections summarized in
Table \ref{tab:SELscales}. Because these are quite small for our
adopted scales, they have only a small impact on the final result.

For each of the scales we have a posterior distribution, $P_k[\DeltaO]$,
describing the likelihood of the values of \DeltaO\ between EHR1 and a reference
nebula for each scale $k$.  We have demonstrated that the relative gas-phase
abundances, \DeltaO, are robust in our adopted SEL scales, and we have no reason
{\em a priori} to weight one scale more than another. Thus for a given reference
nebula, we combine the results from each scale, deriving a posterior
distribution for \DeltaO\ with respect to each of the reference disk \HII\
regions, creating a joint distribution, $P[\DeltaO]$: \begin{displaymath}
  P[\DeltaO] = {\displaystyle \prod_{k} P_k[\DeltaO ]},
\end{displaymath}
which we evaluate numerically.

%%%%%%%%%%%%%%%%%%%%%%%%%
%%
%% Figure 2
%\clearpage
\begin{figure}
\epsscale{0.8}
  \plotone{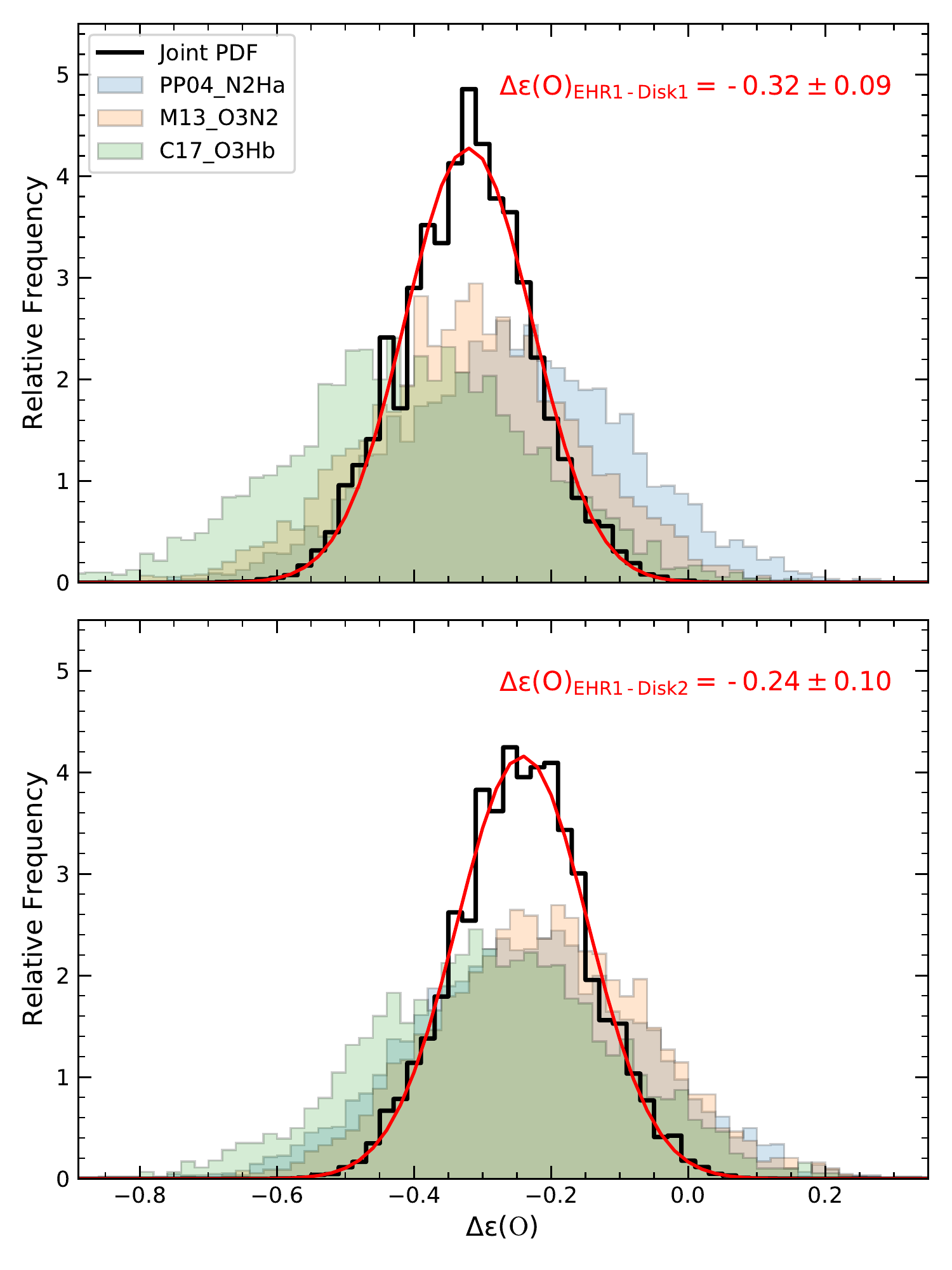}
\caption{Probability density functions for the abundance offset of the
  thick disk \HII\ region \ngc 4013 EHR1 relative to the two reference
  regions in the disk. The top panel shows the results relative to the
  reference nebula disk 1, the bottom relative to disk 2. In both
  panels, the shaded distributions show the results for each of the
  three individual SEL scales used ($ P_k[\DeltaO]$). The black
  histogram shows the joint PDF ($ P[\DeltaO]$) combining the three
  scales, and the red line shows a Gaussian fit to the
  distribution. Our principal results are those relative to the
  reference nebula disk 1, as the data for the disk 2 nebula likely
  suffer from non-uniform slit losses. \label{fig:delta}}
\end{figure}

Figure \ref{fig:delta} shows the individual distributions $P_k[\DeltaO]$ for our
three adopted SEL scales as applied to the EHR1-disk 1 comparison (top panel)
and EHR1-disk 2 comparison (bottom panel). Each panel also shows the joint
distribution $P[\DeltaO]$ describing our final result (all PDFs are normalized).
The individual and combined results are summarized in Table \ref{tab:results}.
Our final estimates of \DeltaO\ for EHR1 compared with each reference region are
based on the joint posterior distribution combining the three adopted SEL
scales. Our favored results are those using only disk 1 as a reference:
$\DeltaO_{\rm EHR1 - disk\, 1} = -0.32 \pm0.09$. As the distributions are close
to normally-distributed, this value is the mean of the distribution based on a
Gaussian fit to the PDF with 68\% confidence errors. If we consider disk 2 as
the reference nebula we find $\DeltaO_{\rm EHR1 - disk\, 2} = -0.24\pm0.10$.
Using the disk nebulae together as a joint reference yields $\DeltaO_{\rm EHR1 -
disk} = -0.28\pm0.06$. We emphasize that EHR1 has an abundance lower than both
disk \HII\ regions in every SEL scale we tested.

\begin{deluxetable*}{lcccc}
  \tabletypesize{\small}
\tablewidth{0pc}
\tablecolumns{5}
\tablecaption{NGC 4013 EHR1 Abundance Summaries \label{tab:results}}
\tablehead{\colhead{Quantity} & \colhead{N2 [PP04]} &
  \colhead{O3N2 [M13]} &
  \colhead{O3 [C17]} & \colhead{Combined}}
\startdata
$\epsO_{\rm EHR1}$      & $\ 8.35\pm0.15$& $\ 8.43\pm0.15$& $\ 8.38\pm0.17$ & $\cdots$ \\
$\DeltaO_{\rm disk\, 1}$ & $-0.26\pm0.17$ & $-0.34\pm0.15$ & $-0.40\pm0.18$ & $-0.32\pm0.09$ \\
$\DeltaO_{\rm disk\, 2}$ & $-0.24\pm0.17$ & $-0.22\pm0.15$ & $-0.28\pm0.18$ & $-0.24\pm0.10$ \\
\enddata
\tablecomments{The abundances and abundance offsets are given for
  the adopted SEL scales. The method and references for the
  calibration are given in the column headers. The \DeltaO\ results
  are also given for a combined or joint result for each reference
  region. The errors are at 68\% confidence. }

\end{deluxetable*}

The ultimate \DeltaO\ values are not overly sensitive to which set of
scales we adopt. If we jointly apply all of the ``reliable'' scales
from Table \ref{tab:SELscales}, we find $\DeltaO_{\rm EHR1 - disk\, 1}
= -0.28 \pm0.05$, and $\DeltaO_{\rm EHR1 - disk\, 2} = -0.22 \pm0.05$,
and jointly $\DeltaO_{\rm EHR1 - disk} = -0.26\pm0.04$. The errors
here are likely underestimated, as we have not accounted for the
correlations when using the same methods in this treatment. However,
the results are instructive in that they demonstrate our specific
choice of adopted scales is not highly biased.

The propagated observational errors are of order 0.01 to 0.03 dex for those
indicators not using \OII; for the \OII -based results, the observational errors
are of order 0.04 to 0.05 dex. The major contributors to the uncertainties in
all of our analyses are the intrinsic dispersions expected when calculating
\DeltaO\ (see Table \ref{tab:SELscales}). While the offsets about the direct
method \DeltaO\ values (used to calculate $\sigma_{\DeltaO}$ in Table
\ref{tab:SELscales}) are not strongly correlated for our adopted scales, we have
not done a full assessment of this for the larger sample of reliable scales.

The {\em absolute} abundances from SEL scales can have significant systematic
uncertainties. However, a representative abundance comes from the N2[$\equiv
I(\NII \, 6584)/I(\Halpha)$] scale of \cite{pettini2004}, which gives
$\epsO_{\rm EHR1} = 8.35\pm0.15$ and $\epsO_{\rm disk\, 1} = 8.63\pm0.15$ (where
the errors are based on the dispersion of this scale about the direct method
abundances from the literature sample discussed above). The full range of median
abundances predicted for EHR1 by our adopted scales are $\epsO_{\rm EHR1} =
8.24$ to 8.49.

%%%%%%%%%%%%%%%%%%%%%%%%%%%%%%%%%%%%%%%%%%%%%%%%%%%%%%%%%%%%%%%%%%%%%%
%% Discussion

\section{The Metallicity of the Thick Disk of \ngc 4013}
\label{sec:discussion}

The thick disk \HII\ region \ngc 4013 EHR1 probes gas at $z = 860$ pc above the
plane of this galaxy that is illuminated and energized by a cluster of young,
hot stars \citepalias[based on the photometric properties and \Halpha\ strength;
see ][]{howk2018b}. This gas has an abundance a factor of $\approx2$ (0.3 dex)
lower than that of the disk of \ngc 4013. The disk gas in \ngc 4013 has an
abundance consistent with that of the local Milky Way disk, which is
approximately ``solar'', $\epsO_\odot = 8.72$ \citep{steffen2015}. The low
metallicity of the thick disk gas illuminated by these OB stars must therefore
either represent (a) low-metallicity gas that has not mixed with material
recently ejected from the disk, or (b) an admixture of recent fountain-ejected
gas with very low metallicity material. It cannot be pure ejected disk gas given
its low abundance.

The interstellar thick disk of \ngc 4013, like many spiral galaxies, shows
strong evidence for the presence of gas lifted from the midplane, perhaps
through an active galactic fountain (although the hydraulic jump mechanism would
lead to similar structure for our purposes). This evidence includes its observed
multiphase structure \citep{rueff2013}, which requires continuous energy input,
the near co-rotation of the thick disk with gas in the plane
\citep{zschaechner2015}, and the presence of significant amounts of dust
\citep[as the dust almost certainly originates in the disk;][]{rueff2013}. And
yet, given the low metallicity of gas at $z = 860$ pc, the thick disk gas in
this galaxy cannot be pure ejected material; it must include a contribution
from lower-metallicity gas.

Gas accreted directly as a cold stream from the IGM would have an extremely low
metallicity, but such matter is likely to be heated to very high temperatures
and held up in the corona of a low-redshift galaxy like \ngc 4013 \citep[e.g.,
][]{birnboim2003, van-de-voort2011, nelson2013, nelson2016}. It is also unlikely
that directly-accreting matter would be deposited into the thick disk at a
projected radial distance of $R \approx 2.5$ kpc from this galaxy's center,
especially if that gas has any initial angular momentum relative to the disk.
(EHR1 has a velocity similar to that of the diffuse ionized and warm neutral gas
at its position, so its rotation is not at odds with that of galactic
fountain-driven gas.) The same argument makes it unlikely the low-metallicity
nebula \ngc 4013 EHR1 is tracing gas from an accreted dwarf satellite \cite[the
``bulls-eye'' problem of ][]{peek2009}. Though there is evidence for a major
merger several Gyr ago \citep{wang2015} based on \ngc 4013's prodigious warp
\citep{bottema1996, zschaechner2015} and recently-discovered stellar stream
\citep{martinez-delgado2009}, the probability of any gas from that merger
landing within $R\approx 2.5$ kpc of the center of \ngc 4013 is small. (And
finding it there would be even less likely since thick disk gas is returned to
the plane in $\sim100$ Myr. Thus any material from an accreted dwarf would not
find itself $z \approx 1$ kpc above the midplane for very long.)

More likely is that any newly-acquired, low-metallicity matter is subsumed into
the hot, massive corona of the galaxy and accreted onto the disk at a later
time.  The coronal matter must be cooled to bring it to the low \z -heights and
cool, dense conditions we observe. If the corona does have a low metallicity (in
this case, $\la 1/2$ solar), cooling it may be problematic, as the efficacy of
thermal instabilities in low-metallicity coronal gas is debated
\citep{binney2009}. However, such cooling may be made possible by
fountain-induced accretion\footnote{Although, given how the mechanism works,
metal rich gas ejected through any mechanism (e.g., through a hydraulic jump)
should provide similar results.} \citep{marinacci2010, marinacci2012,
marasco2012}.  These models postulate that cold, metal-rich gas ejected from the
disk can mix with hot, metal-poor coronal gas, enhancing its cooling, which
causes the coronal gas to accrete onto the disk. However, our measured $\DeltaO
\approx -0.3$ dex between the thick and thin disk material in \ngc 4013 is
larger than these models are currently able to reproduce. If we assume the
coronal material is pristine with no metals (an extreme model), equal amounts of
coronal and fountain material would need to be mixed in order to produce the
observed metallicity offset between EHR1 and the disk.  For a higher-metallicity
corona, a larger fraction of cooled coronal material is required.  The
fountain-driven accretion models predict that condensed coronal material will
only contribute $\approx10-20\%$ of the total mass of extraplanar gas
\citep{marasco2012}, although such cooling can be more effective in
lower-temperature coronae \citep{armillotta2016}. Very recent simulations
\citep{gritton2017} of metal-poor clouds flowing through a {\em metal-rich}
corona show promise for condensing a larger amount of material, but it is not
yet clear if they translate to the situation appropriate for \ngc 4013.

One could imagine the low metallicity of the gas in EHR1 could be the result of
the collision of a low-metallicity HVC with the disk \citep[e.g.,]{franco1988,
lepine1994, lockman2008, park2016}. In this case, the HVC material itself would
provide the low-metallicity gas. This is somewhat different from direct
accretion if HVCs represent material condensed from the hot, metal-poor corona
and thus have relatively low angular momentum. It would still be unusual to
observe this process directly given the timescales involved, but not unheard of
\citep[e.g.,][]{lockman2008}. Such a collision could be partially responsible
for triggering the formation of dense clouds in the thick disk and ultimately
the stars that power \ngc 4013 EHR1 \citep[e.g.,][]{martos1998}. The nature and
formation of the stars underlying EHR1 are discussed in \citetalias{howk2018b}.

\section{Vertical Abundance Gradients in Spiral Galaxies}
\label{sec:vertgradients}

\subsection{The Literature Sample of Thick Disk \HII\ Regions}

Previous metallicity determinations are available for two extraplanar nebulae in
the dwarf galaxy \ngc 55 \citep{tullmann2003} and for three nebulae in two
massive spirals undergoing stripping \citep{stein2017}.  We have re-derived
\DeltaO for each of these, using the line intensities provided in the original
references and following the approach described in \S \ref{sec:montecarlo}. Our
results comparing each extraplanar nebula in these galaxies with their
respective disks are given in Table \ref{tab:offsets}.

\begin{deluxetable*}{lcccc}
  \tabletypesize{\footnotesize}
  \tablewidth{0pc}
  \tablecolumns{5}
  \tablecaption{Extraplanar \HII\ Region Abundance Offsets
    \label{tab:offsets}}
  \tablehead{\colhead{Galaxy} & \colhead{Nebula ID} &
    \colhead{\z\ (pc)} & \colhead{\DeltaO} & \colhead{Reference}}
  \startdata
  NGC~4013  & 1 & 0.9  & $-0.32\pm 0.09$ & This work \\
  NGC~55    & 1 & 1.1  & $+0.24\pm 0.12$ & \cite{tullmann2003} \\
  NGC~55    & 2 & 2.2  & $+0.29\pm 0.12$ & \cite{tullmann2003} \\
  NGC~3628  & 2 & 2.8  & $-0.17\pm 0.10$ & \cite{stein2017} \\
  NGC~3628  & 3 & 3.0  & $-0.05\pm 0.11$ & \cite{stein2017} \\
  NGC~4522  & 1 & 1.4  & $-0.03\pm 0.07$ & \cite{stein2017} \\
  \enddata
  \tablecomments{The nebular IDs are those reported in the original
    references. All abundance offsets \DeltaO\ are derived anew using
    the intensities reported in the original references. Where more than
    one disk reference region is available (for \ngc 4013 and \ngc
    4522), the values for \DeltaO\ represent the joint distribution. The
    medians of the distributions are given along with errors
    representing 68\% confidence.}
\end{deluxetable*}

For the extraplanar \HII\ regions recently presented by \cite{stein2017} in two
massive spiral galaxies, our calculations of \DeltaO\ are in reasonable
agreement with their determinations (within 0.1 dex, the typical error in our
assessments). These nebulae in \ngc 3628 and \ngc 4522 are plausibly associated
with tidally- or ram pressure-stripped gas about these group (\ngc 3628) and
cluster galaxies (\ngc 4522).

For the \HII\ regions seen above the plane of \ngc 55 we find significantly
different values for the absolute abundances than the original publication
\citep{tullmann2003}, and our values imply a different sign for \DeltaO . Our
analyses of the \HII\ regions in \ngc 55 yield higher abundances in the two
extraplanar \HII\ regions than the single disk region in the \cite{tullmann2003}
spectroscopy. Our assessments of the abundance of the reference disk \HII\
region range from $\epsO = 8.1$ -- 8.3.  For comparison \cite{tullmann2003}
calculated $\epsO = 8.05\pm0.10$ using the direct method for the disk \HII\
region. \cite{kudritzki2016} derived $\epsO \approx 8.32\pm0.03$ as the central
abundance of \ngc 55 based on measurements of the atmospheres of blue
supergiants. Thus the absolute abundance we derive for the disk reference is
consistent with the range of likely values.  Our determinations of the
abundances for the extraplanar \HII\ regions cover $8.1 \la \epsO \la 8.7$.  For
the extraplanar regions, every scale denoted ``reliable'' in Table
\ref{tab:SELscales} yields a higher abundance than the disk \HII\ region when
using the same scale.

The difference in our results for \ngc 55 compared with those of
\cite{tullmann2003} are understandable given the advances in understanding SEL
methods and their limitations since the original publication.
\citeauthor{tullmann2003} used the M91 \citep{mcgaugh1991} calibration of
\Rtwothree\ to derive absolute abundances.  When using \Rtwothree, these nebulae
are in a region of some ambiguity as to the correct choice of abundance from
this indicator \citep{kewley2008}.  At the time that \citeauthor{tullmann2003}
derived their metallicities, the tools for assessing which branch to adopt for
this double-valued indicator were less robust.

Extraplanar \HII\ regions in the two massive spirals \ngc 3628 and \ngc 4522
show modest or no abundance offsets relative to their disks. Our results for
these galaxies are consistent with the conclusions reached by \cite{stein2017}.
In particular, the nebula at $z\approx 1.4$ from the midplane of \ngc 4522 is
projected onto extraplanar \HI\ identified as matter being stripped from the
disk via the ram pressure interaction with the intracluster medium of the Virgo
cluster in the maps of \cite{kenney2004}. \cite{stein2017} note the extraplanar
\HII\ regions in \ngc 3628 are projected onto an \HI\ filament seen in the maps
of \cite{wilding1993} that has been identified as a tidal feature due to the
interaction of this galaxy with its neighbor \ngc 3627. Thus, the extraplanar
\HII\ regions may plausibly arise in gas stripped from both of the galaxies, and
they represent fundamentally different phenomena than the extraplanar \HII\
regions in \ngc 55 and \ngc 4013.

The spiral galaxy \ngc 4522 resides in the Virgo cluster; direct imaging shows
spectacular evidence for ram pressure-stripping of gas from the disk
\citep{abramson2014}. The extraplanar \HII\ region measured in this galaxy is
associated with this material that seems to clearly have an origin in the disk
of that galaxy \citep{stein2017}. We derive a joint abundance offset between the
two disk \HII\ regions observed and the single extraplanar \HII\ region of
$\DeltaO = -0.03 \pm0.07$ (68\% confidence), consistent with no offset. This is
as expected for material recently stripped from the disk.

In the case of \ngc 3628 (part of the Leo Triplet of galaxies), two extraplanar
\HII\ regions are seen at $z \approx3$ kpc in close proximity to one another.
\citeauthor{stein2017} note that these \HII\ regions are likely associated with
a filament of \HI\ at similar velocities that likely represents tidal material
due to an interaction of \ngc 3628 with its neighbor \ngc 3627
\citep{wilding1993}. There are hints that the filament may have arisen in the
outer parts of \ngc 3628 \citep{stein2017}, but they seem inconclusive. Compared
with the single disk \HII\ region measured in that galaxy, we find offsets of
$\DeltaO = -0.17 \pm 0.10$ and $-0.05\pm0.11$ (68\% confidence). If one assumes
a single abundance offset is appropriate, the joint distribution implies
$\DeltaO = -0.12 \pm 0.07$. There is a hint that the abundances of these \HII\
regions may have a slightly lower abundance than the disk, but the hint is not
robust at the 95\% level.

While the extraplanar regions tracing stripped material have abundances
consistent with the disks of those galaxies, the extraplanar nebulae in the
dwarf galaxy \ngc 55 have higher abundances than the disk. The extraplanar \HII\
regions in \ngc 55 seem to be connected with supershells produced by vigorous
star formation-driven feedback \cite{tullmann2003}. Assuming the
recently-determined distance to \ngc 55 \citep[$D = 2.34$
Mpc;][]{kudritzki2016}, these are projected $z\approx 1.1$ and 2.2 kpc from the
midplane (compared with the previous estimate of $z\approx0.8$ and 1.5 kpc).
\cite{ferguson1996} first identified these \HII\ regions based on \Halpha\
imaging. The lower \HII\ region is found within a large supershell on the
northern side of the galaxy, while the upper one is found at the end of a long
($\sim1.5$ kpc) filament of \Halpha\ emission on the southern side of the
galaxy.

Both \cite{ferguson1996} and \cite{tullmann2003} have speculated whether the
processes that produce these large \Halpha -emitting structures in \ngc 55 may
have played a role in triggering the formation of stars in the thick disk of
this galaxy. This is consistent with our \DeltaO\ determinations: if
feedback-driven outflows are the ultimate source of the gas emitting in these
nebulae, one may expect that gas to be more metal rich than the disk if there
has been little mixing with more metal-poor material.

%% However, it is not clear that the stars energizing these nebulae
%% need to have formed {\em in situ}. The ionizing sources seem to be
%% consistent with single later-type OB stars ($\sim$B0 V stars). Such
%% stars, with lifetimes $\sim7.5$ Myr, could reach $z \approx 2.2$
%% kpc within their lifetimes for velocities $v_z \la 300$
%% \kms. \citep[This is in contrast to the case for \ngc 4013, where
%% the illumination of the \HII\ region is provided by multiple
%% stars;][]{howk2018b}.  \cite{tullmann2003} argue that
%% hydrodynamical drag on the \HII\ region material would prevent it
%% from moving at such a large velocity. In that we agree. However, it
%% is only the stars that need to be ejected at that velocity. They
%% will ionize whatever gas happens to be in their vicinity at a given
%% time. The mechanisms often invoked for the ejection of runaway
%% stars, e.g., by supernova explosions or black hole encounters, are
%% expected to produce single stars with velocities comparable to or
%% higher than this \citep{gvaramadze2009}. Hence, one cannot conclude
%% that these stars must have formed {\em in situ} above the disk of
%% \ngc 55.

\subsection{Vertical Abundance Gradients in Context}

\begin{figure*}
\plotone{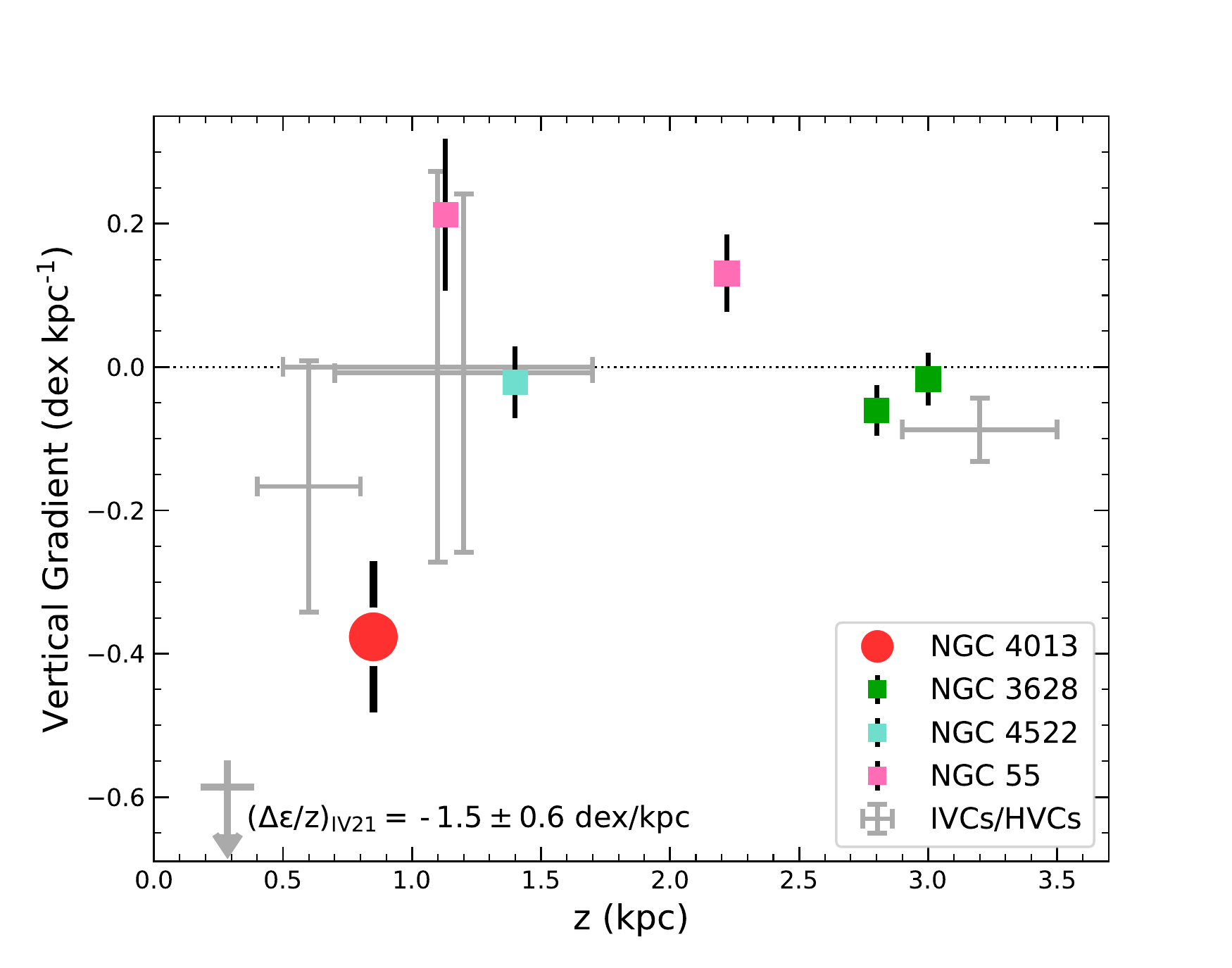}
\caption{Abundance gradients, $\DeltaO / z$, implied by the existing sample of
extraplanar \HII\ regions \citep{tullmann2003, stein2017} as a function of
height from the midplane. The \DeltaO\ values have been derived for all regions
as for EHR1; thus the values used here may differ from those in the original
references. In particular, the extraplanar \HII\ regions in NGC 55 have a
positive offset relative to the disk \citep[contrary to earlier results
][]{tullmann2003}. The systems shown by squares (in NGC~3628 and NGC~4522)
likely arise in stripped gas, perhaps from the host galaxy. The extraplanar
nebulae in NGC~55 are likely associated with gas participating in a galactic
fountain circulation. The high-\z\ nebula in NGC~4013 may be a mixture of
fountain gas and gas much lower metallicity. Also shown are results from Milky
Way IVCs \citep{wakker2001,richter2001,sembach2004,hernandez2013} and one HVC
\cite[the Smith Cloud;][]{fox2016}. Other HVCs with abundance measurements are
at much larger \z. The IVC IV21 \citep{hernandez2013} resides at a value of
$-1.5\pm0.6$ on the vertical scale (with $\Delta \epsilon \approx -0.4$ and $|z|
\approx 285$ pc); we place it at an arbitrary position vertically in this
figure. \label{fig:gradients}}
\end{figure*}

We compare the implied gradient in abundances for the complete
literature sample of extraplanar \HII\ region abundance determinations
\citep{tullmann2003,stein2017} in Figure \ref{fig:gradients}. We also
include results from the Milky Way intermediate and high velocity
clouds \citep[IVCs and HVCs;][]{wakker2001, richter2001, sembach2004,
  hernandez2013, fox2016}. In this context, we define the gradient
simply as $\DeltaO / z$ for each \HII\ region or cloud. The full
sample of extraplanar \HII\ region abundances shown in Figure
\ref{fig:gradients} paint a mixed picture of vertical abundance
gradients in galaxies. Given the small sample size, and particularly
given the variety of origins plausibly attributable to each, it is not
clear what overarching conclusions can yet be drawn.

%%%%%%%%%%%%%%%%%%%%%%%%%
%%
%% Figure 4

The factor of two abundance offset between the extraplanar \HII\ region EHR1 and
the disk of \ngc 4013 is reminiscent of the difference seen between some
Galactic HVCs and the Milky Way's disk \citep{wakker2001, barentine2013}.
However, the HVCs tend to be much further from the Galactic disk than $z\sim1$
kpc \citep{lehner2011a, lehner2012}, giving smaller gradients in Figure
\ref{fig:gradients}. The Galactic IVCs are a better match to the height probed
by \ngc 4013 EHR1 \citep{wakker2001}, and they are the closest Milky Way analogs
to the massive, dusty clouds seen in direct imaging of edge-on galaxies
\citep{howk1999, howk2000, rueff2013}.  Only one IVC, IV21 at $|z| \approx 285$
pc from the plane of the Milky Way, has a low metallicity (measured principally
using S) with $\epsilon ({\rm S}) = -0.43\pm0.12$ \citep{hernandez2013}. All of
the other clouds within the first few kpc are consistent with zero metallicity
gradient. The metallicity gradients in the thick disk of the Milky Way are
generally quite small within $z \sim 1 - 3$ kpc.

The gradients represented by EHR1 in \ngc 4013 and IV21 in the Milky Way are the
largest observed, in part because they are at small $z$. And, indeed, this is in
part what makes both examples so interesting. Figure \ref{fig:gradients}
demonstrates that the scales for metallicity changes can be much smaller in the
vertical than the radial directions in spiral galaxies, with the caveat that we
cannot yet demonstrate there are true gradients as opposed to random mixtures of
different metallicity gas. There are several other galaxies with extraplanar
\HII\ regions suitable for spectroscopy such as that described here, and these
may help clarify the picture.

The interstellar thick disks of galaxies are complex admixtures of gas having
different origins. Our results suggest large metallicity inhomogeneities exist
on scales of only 100s of pc (certainly between the mid-plane and the thick
disk, but even within the thick disk). In addition, the low metallicity of EHR1
relative to the disk of \ngc 4013 and of IV21 relative to the Milky Way's disk
\citep{hernandez2013} suggest the fueling of galactic disks by low-metallicity
gas is robust at low-redshift, which has corresponding implications for chemical
evolution calculations. The observational questions moving forward will be
whether such low-metallicity inclusions are common in spiral galaxies (e.g.,
Figure \ref{fig:gradients}) and whether the metallicity variations are organized
or stochastic.

Ultimately our favored model for explaining the low abundance seen in the thick
disk of \ngc 4013 is some form of induced accretion \citep{marinacci2010,
marasco2012, fraternali2017}. However, if such large abundance offsets are
common, more work will need to be done to understand if the cooling efficiency
of that model can be increased.

\section{Summary}
\label{sec:summary}

We have presented optical spectroscopy of an \HII\ region in the thick disk of
the nearby spiral galaxy \ngc 4013. The nature of this \HII\ region and its
underlying stars is discussed in \citetalias{howk2018b}. Here we consider the
relative abundance of this nebula compared with the disk in order to understand
the nature of the thick disk in spiral galaxies.  Our principle conclusions are
as follows.

\begin{enumerate}

\item The emission line spectrum of \ngc 4013 EHR1 is generally consistent with
those of \HII\ regions in the disks of spiral galaxies. Thus, it is a relatively
luminous \HII\ region located $z = 860$ pc above the plane of this spiral
galaxy. The \halpha\ luminosity implies the presence of the equivalent of $\sim
6$ O7 V stars \citepalias[see][]{howk2018b}.

%% \item Carefully-chosen SEL abundance scales can give accurate measures
%%   of the relative abundances, \DeltaO, between \HII\ regions.

\item The abundance of EHR1 is a factor of $\approx2$ below that of the disk of
\ngc 4013 (as assessed using two \HII\ regions in the midplane). Our preferred
value for the abundance offset is $\DeltaO_{\rm EHR1 - disk 1} = -0.32 \pm
0.09$.

\item The low abundance of EHR1 implies significant amounts of low-metallicity
gas have been mixed into the thick disk of \ngc 4013. This is particularly
interesting given the presence of large amounts of material expelled from the
disk as part of an active galactic fountain or hydraulic jump in this galaxy.

\item A small sample of measurements of vertical abundance gradients from
extraplanar \HII\ regions and extraplanar clouds in the Milky Way does not yet
show consistent trends. There are examples of positive, negative, and no
abundance gradients with $z$-height. It is not clear if the gradients that are
seen are organized or stochastic.

\end{enumerate}

\acknowledgements

JCH recognizes the hospitality of the Instituto de Astrof\'{i}sica,
Pontificia Universidad Cat\'{o}lica de Chile during the writing of
this work. Portions of this work has been supported by NASA through
grant NNX10AE87G as well as the NSF through grant AST-1212012. It
makes use of data collected at the LBT, an international collaboration
among institutions in the United States, Italy and Germany. LBT
Corporation partners are: The University of Arizona on behalf of the
Arizona Board of Regents; Istituto Nazionale di Astrofisica, Italy;
LBT Beteiligungsgesellschaft, Germany, representing the Max-Planck
Society, The Leibniz Institute for Astrophysics Potsdam, and
Heidelberg University; The Ohio State University, and The Research
Corporation, on behalf of The University of Notre Dame, University of
Minnesota and University of Virginia.

\software{Astropy \citep{price-whelan2018},
  Matplotlib \citep{hunter2007}, \pymcz\ \citep{bianco2016}}

\facilities{LBT(MODS), HST(WFPC2), WIYN}

%%%%%%%%%%%%%%%%%%%%%%%%%%%%%%%%%%%%%%%%%%%%%%%%%%%%%%%%%%%%%%%%%%%%%%
%%
%% REFERENCES

\bibliography{edge_on_galaxies}

%%%%%%%%%%%%%%%%%%%%%%%%%%%%%%%%%%%%%%%%%%%%%%%%%%%%%%%%%%%%%%%%%%%%%%
%%
%% FIGURES

%%%%%%%%%%%%%%%%%%%%%%%%%%%%%%%%%%%%%%%%%%%%%%%%%%%%%%%%%%%%%%%%%%%%%%
%%
%% TABLES

\clearpage

        %%%%%%%%%%%%%%%%%%%%%%%%%%%%%%%%%%%%%%%%%%%%%%%%%%%%%%%%%%%%%%%%%%%%%%

\end{document}